\documentclass{aa}
\usepackage{times}
\usepackage{paralist}
\usepackage{float}
\usepackage{array}
\usepackage{tabularx}
\usepackage{changepage}
\usepackage{color}
\usepackage[latin1]{inputenc}
\usepackage{graphicx}
\usepackage{natbib}
\usepackage{mathabx}
\usepackage{multirow}
\usepackage{txfonts}
\usepackage{bm}

\def\degr{\hbox{$^\circ$}}
\def\arcmin{\hbox{$^\prime$}}
\def\arcsec{\hbox{$^{\prime\prime}$}}
\begin{document}
   \title{A new method to suppress the bias in polarised intensity}


   \author{Peter M\"uller\inst{1}, Rainer Beck\inst{1}, and Marita Krause\inst{1} }

   \institute{\inst{1} Max-Planck-Institut f\"ur Radioastronomie,
              Auf dem H\"ugel 69, 53121 Bonn, Germany\\
              \email{peter@mpifr-bonn.mpg.de}
              }

   \date{Received 2016 Aug 23; accepted 2016 Dec 18}

\titlerunning{Suppressing the bias in polarised intensity}

\abstract{Computing polarised intensities from noisy data in Stokes $U$ and $Q$ suffers from a positive bias that
should be suppressed.}
{To develop a correction method that, when applied to maps, should provide a distribution of polarised intensity that closely
follows the signal from the source.}
{We propose a new method to suppress the bias by estimating the polarisation angle of the source signal in a
noisy environment with help of a modified median filter. We then determine the polarised intensity, including the noise, by projection of the 
observed values of Stokes $U$ and $Q$ onto the direction of this polarisation angle.}
{We show that our new method represents the true signal very well. If the noise distribution in the maps of $U$ and $Q$
is Gaussian, then in the corrected map of polarised intensity it is also Gaussian. Smoothing to larger Gaussian
beamsizes, to improve the signal-to-noise ratio, can be done directly with our method in the map of the polarised intensity.
Our method also works in case of non-Gaussian noise distributions.}
{The maps of the corrected polarised intensities and polarisation angles are reliable even in regions with weak signals and provide
integrated flux densities and degrees of polarisation without the cumulative effect of the bias, which especially affects faint sources. Features
at low intensity levels like 'depolarisation canals' are smoother than in the maps using the previous methods, which has broader implications, for example on the
interpretation of interstellar turbulence.}


   \maketitle
%

\section{Introduction}

Linearly polarised emission is a powerful tool in astrophysics. Scattering of light generates optical polarisation that can constrain the geometry of reflection 
nebulae (e.g. Scarrott et al. 1986). Photons can also be polarised by scattering or extinction at elongated dust grains aligned in interstellar magnetic fields 
(e.g. Hoang \& Lazarian 2014), which allows mapping these magnetic fields in the Milky Way (Fosalba et al. 2002) and in the Small Magellanic Cloud (Gomes et al. 2015). 
Elongated dust grains emit polarised emission at sub-mm wavelengths, which is useful to study the magnetic fields in molecular clouds 
(e.g. Tang et al. 2009, Pillai et al. 2015) or the halos of galaxies (e.g. Greaves et al. 2000). Recently, the PLANCK mission provided all-sky dust polarisation 
maps revealing large-scale magnetic fields in the Milky Way (Planck Collaboration 2015).

Synchrotron emission is up to 75\% linearly polarised, with its B-vector intrinsically parallel to the magnetic field. Optical synchrotron emission allows the 
investigation of magnetic fields of jets emerging from galactic nuclei (e.g. Perlman et al. 2011) and radio synchrotron emission the investigation of magnetic fields 
in the Milky Way (Wolleben et al. 2006), other spiral galaxies (e.g. Beck 2016), and radio galaxies (e.g. Laing \& Bridle 2014).

Linearly polarised emission is usually described in terms of the Stokes parameters $U$ and $Q$, defined as $U=P\,sin(2\,\chi)$ and $Q=P\,cos(2\,\chi)$, where $P$ is the polarised intensity and $\chi$ is the polarisation angle. If the receiving system delivers orthogonally polarised signals with amplitudes $X$ and $Y$, $U$ and $Q$ are computed as $U=2\,X\,Y\,cos\,\delta$ (where $\delta$ is the phase between $X$ and $Y$) and $Q=X^2-Y^2$. If circularly polarised signals with amplitudes $L$ and $R$ are delivered by the receiving system, $U=2\,L\,R\,sin\,\delta$ and $Q=2\,L\,R\,cos\,\delta$, where $\delta$ is the phase between $L$ and $R$.
Low-frequency radio telescopes consisting of dipoles, like the LOw Frequency ARray (LOFAR), deliver projections of $X$ and $Y$ signals, whereas most higher-frequency 
radio telescopes use correlating devices to transform $L$ and $R$ signals into $U$ and $Q$.

Measuring polarised signals suffers from a fundamental problem.
In addition to the true polarised signal $P_T$ of a source, root mean squared (rms) noise obtained from the receiving system is also detected.
If we use the standard formula to calculate $\hat{P}$ from the measured data $\hat{U}$ and $\hat{Q}$, $\hat{P}$ can be expressed by the polarised components
of the source $U_T$ and $Q_T$ and their noise contributions $N_U$ and $N_Q$:
\begin{equation}
 \begin{array}{lcl}
 \hat{P} &=& \sqrt{\hat{U}^2 + \hat{Q}^2} \\
        &=& \sqrt{(U_T+N_U)^2+(Q_T+N_Q)^2} \, .
 \end{array}
\end{equation}

The noise always delivers a positive bias to the true polarised intensity $P_T$ that cannot be separated out for small
signal-to-noise ratios $s$ ($s = \hat{P}/\sigma$, where $\sigma$ is the rms noise in the maps of $\hat{U}$ and $\hat{Q}$).
Though this could be overcome by clipping maps of polarised intensity below a certain value, the bias
accumulates by integration of polarised intensities and, if it cannot be effectively suppressed, prevents the determination of reliable flux densities in polarisation. 
Further, due to the noise bias, the distribution of $\hat{P}$ is Ricean.
As a result, further data processing, for example
smoothing directly to a larger beam size (this is usually done instead by smoothing the $\hat{U}$ and $\hat{Q}$ maps, which is subject to depolarisation).

The distribution of $\hat{P}$ is Ricean for small $s$ and becomes
Gaussian for large $s$. Vinokur (1965, Eqs. (81) and (82)) showed that a
separation of signal $P_T$ and standard deviation of the noise distribution $\sigma$ is possible only for large $s$ and the expectation value is:
\begin{equation}
\langle P \rangle = P_T + \sigma^2 / \,(2 \,P_T) \, ,
\end{equation}
while for small $s$
\begin{equation}
\langle P \rangle = \sqrt{\pi/2} \, \sigma \, (1 + s^2 / 4) \, .
\end{equation}

Several other methods have been developed to correct for the noise bias.
The most widely used method is that of Wardle \& Kronberg (1974) who proposed the following bias correction
for $\hat{P}$ maps:
\begin{equation}
P^* = \sqrt{\hat{P}^2 - \sigma^2}
\end{equation}
where $\sigma$ is the rms standard deviation of the noise distributions in $\hat{U}$ and $\hat{Q}$.

An astrophysical interpretation requires that a map of bias-corrected $P^*$ has a base level of about zero in regions containing only noise.
This requires introduction of negative values of $P^*$ at locations where the true signal $P$ is very weak:
\begin{equation}
\begin{array}{lcl}
P^* &=& \sqrt{\hat{P}^2 - (C\, \sigma)^2} \ \  (\mathrm{if} \,\, \hat{P}^2 \ge (C\, \sigma)^2) \, ,\\
P^* &=& - \sqrt{- \hat{P}^2 + (C\, \sigma)^2} \ \  (\mathrm{if} \,\, \hat{P}^2 < (C\, \sigma)^2) \, .
\end{array}
\end{equation}
The smallest possible value of $P^* = -C\,\sigma$ occurs for $\hat{U} + \hat{Q} = 0$.
The factor $C$ was introduced later to adjust the bias correction: $C=1.2$ is used in the NOD2 software
package (Haslam 1974, Andernach 1985) and AIPS (Greisen 2003) uses $C=1.253$ in the option POLC of the task
COMB. If $\hat{P}$ is determined from multi-channel data with help of rotation measure (RM) synthesis, the
additional uncertainty in RM requires that $C$ is increased to $1.5$ (George et al. 2012).

The AIPS task POLCO uses the maximum likelihood correction introduced by Killeen et al. (1986) for large signal-to-noise ratios ($s>2$) and the Wardle \& Kronberg 
method for smaller $s$.

Simmons \& Stewart (1985) discussed various estimators of the true signal $P$, including a maximum likelihood, a median estimator, and the Wardle \& Kronberg estimator. 
All estimators agree asymptotically for large $s$. All estimators
yield a positive residual bias (relative to $P$) for $s<1$, but a negative for $1<s<4$ and hence are not a good
representation of the true signal for $s<4$. Montier et al. (2015) introduced a Bayesian estimator that provides a very low relative bias for  $s<1.2$, allowing 
reliable estimates of $P$ in regions with low $s$, while the maximum likelihood estimator performs best for $s>2$. However, none of the methods works well at 
small $s$ and large $s$ simultaneously.

If the polarisation angle is perfectly well known, the estimator proposed by Vidal et al. (2016) is able to completely correct the polarisation bias. This method can 
be applied to regions where the polarisation angle is expected to be constant, such as for large-scale magnetic fields as observed in sub-mm dust emission with WMAP and 
PLANCK.

The new method proposed here works for all values of $s$ and does not need prior knowledge of the polarisation angle.
All methods to compute $P^*$ maps require that the base levels in the maps of $\hat{U}$ and $\hat{Q}$ are about zero
(i.e. their mean values are smaller than about 20\% of the rms noise values) in regions without sources. Baseline shifts may remain even after
processing and combining single dish maps.

\section{The method}

When detecting astronomical signals at the telescope, statistical receiver noise is added to the $U$ and $Q$ signals.
Any linear and non-linear combination of the noisy $\hat{U}$ and $\hat{Q}$ signals should take into account the probability density of the noise
distribution in the resulting data. If we could mathematically express noise and the signal as a linear combination of both, it is possible to deal with the noise
separately. This is not the case if we compute the square root of $U^2$ plus $Q^2$.

\begin{figure}[h]
 \centering
 \includegraphics[scale=0.35, clip=true,trim=20pt 10pt 0pt 0pt]{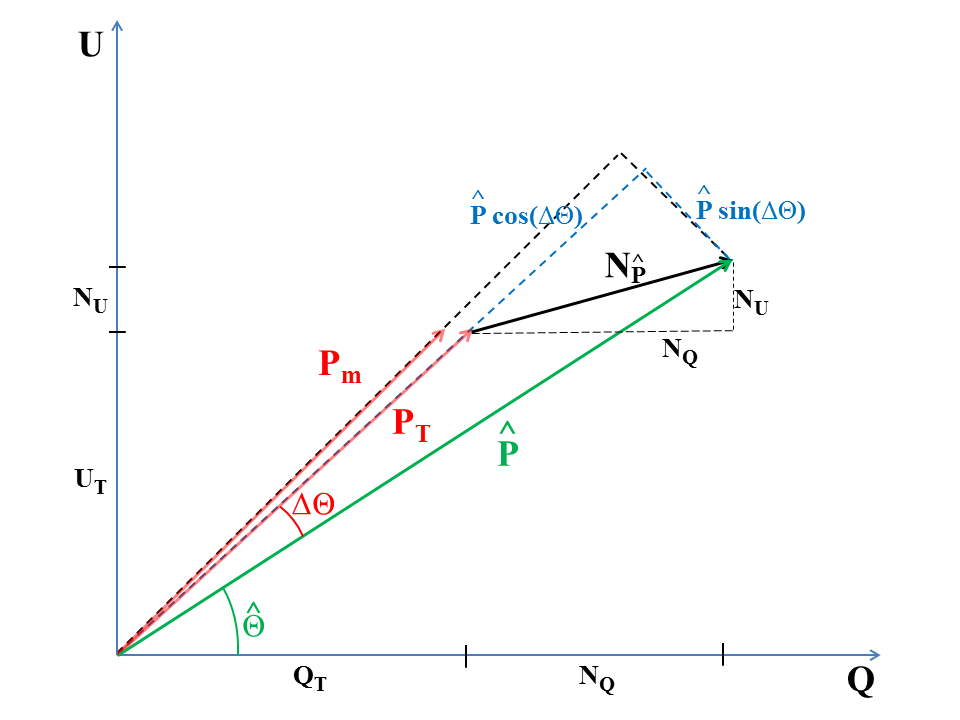}
 \caption{Geometrical sketch showing the noise components of the polarisation vectors. The noise vector $N_{\hat{P}}$ can be split in two different
 orthogonal components. Statistically, all $N_U$ and $N_Q$ contain the measured noise distributions, so the other two projected noise components calculated by the 
 projection of $\hat{P}$ onto $P_T$, respectively $P_m$, contain the same distribution.} 
 \label{Hist_fromPI}
\end{figure}

We express the polarised intensity $P$ in a different way.
For each pixel, we convert $U$ and $Q$ to polar coordinates and define the geometrical angle $\theta = tan^{-1}(U/Q)$ (twice the polarisation angle), so that:
\begin{equation}
 \begin{array}{lcl}
  U &=& P\,sin(\theta) \, , \\
  Q &=& P\,cos(\theta) \, , \\
  P &=& U\,sin(\theta)+ Q\,cos(\theta) \, .
 \end{array}
\end{equation}
It is then possible to separate the true signals $U_T$, $Q_T$ and $P_T$ from their noise contributions $N_U$ and $N_Q$.
The noise contribution of $N_P$ depends on the method applied and is given by the projection of $N_U$ and $N_Q$ if the true angle $\theta_T$ is known:
\begin{equation}
 \begin{array}{rcl}
  P_T+N_P &=& (U_T+N_U)\,sin(\theta_T) + (Q_T+N_Q)\,cos(\theta_T) \, ,\\
  \theta_T &=& tan^{-1}(U_T/Q_T) \, ,
 \label{pinew}
 \end{array}
\end{equation}
where $\theta_T$ is the true angle in the polar representation of the noise-free signals $U_T$ and $Q_T$, as illustrated in Fig.~1.
The observed quantities are defined as $\hat{U} = U_T + N_U$, $\hat{Q} = Q_T + N_Q$, so that 
$\hat{\theta} = tan^{-1}(\hat{U}/\hat{Q})$ and $\hat{P} = \sqrt{\hat{U}^2 + \hat{Q}^2\ }$.
Equation~\ref{pinew} can also be expressed as the projection of the vector $\hat{P}$ onto the direction of the true $P_T$ as follows:
\begin{equation}
 P_T + N_P = \hat{P}\ cos(\theta_T - \hat{\theta}) = \hat{P}\ cos(\Delta\theta) \, ,
 \label{piproj}
\end{equation}
which can easily be proven by expanding $cos(\theta_T - \hat{\theta})$ and comparing with $\hat{U}\,sin(\theta_T) + \hat{Q}\,cos(\theta_T)$.

Application of this equation requires the knowledge of the true angle $\theta_T = tan^{-1}(U_T/Q_T)$ of the signal.
This is the crucial step for this method.
For an individual pair of measured $\hat{U}$ and $\hat{Q}$ it is impossible to know their noise contributions. For a number $n$ of single
observations on a source, or a map of $n$ pixels around the source, we reduce the angle error $\theta_n - \theta_T$ significantly by applying an averaging
filter to adjacent pixels. We choose the median filter, because it reduces the bias approximately the square root of the number
of the averaging pixels, and gives a reliable value for sudden changes of the polarisation angles at the adjacent pixels. However, a problem occurs with the 
discontinuity of the angle $\theta$ from -$\pi$ to $\pi$. Therefore the components $x~=~cos(\theta)$ and $y~=~sin(theta)$ are taken for the median filtering 
and the $arctan2(y_m, x_m)$ is used to calculate the angle $\theta$ from the medians of $x$ and $y$.
The adjacent pixels are not independent from each other due to the smoothing of the telescope beam.
The pixel size is usually chosen to be one third of the beam size, for which
a median filter size of $5\times5$ pixels is a good choice. In this case the angle error $\Delta{\theta_m}$ of the median value $\theta_m$ is
reduced by a factor of approximately five.

We have to take into account that the angle $\hat{\theta} = tan^{-1}(\hat{U} / \hat{Q})$ represents the bias at the centre position of the filter box, which leads
to an undercorrection of the bias.
Therefore we remove this angle at the centre position of the filter box to suppress this bias again at noise level ('modified median filter').
However, disregarding this pixel leads to an overcorrection. In Fig.~2 we used the weighted average (1:2) of the median and the modified median of the filter box, which is an empirical
choice to optimise the bias correction.

In case of overlapping sources with different polarisation angles, there is a depolarisation effect
due to the beam smoothing. The application of the modified median filter may increase the error
in $P^*$. We note that the averaging of the angle does not affect the angular resolution of the $P^*$ map, other than smoothing the $\hat{U}$ and $\hat{Q}$ maps.

From the observations, we get $\hat{U}=U_T+N_U$ and $\hat{Q}=Q_T+N_Q$ and the modified median filtered angle $\theta_m$, and we can separate $P_m$ and its noise
contribution $N_{P_m}$.
The term $P_m + N_{P_m}$ is the resulting polarised intensity $P^*$ determined with our new method, calculated with the angle $\theta_m$ that should be very close to
the true angle $\theta_T$:\

\begin{equation}
\begin{array}{lcl}
P^* &=& P_m + N_{P_m} = (U_T+N_U)\,sin(\theta_m) + (Q_T+N_Q)\,cos(\theta_m) \\
    &=& \hat{U}\,sin(\theta_m) + \hat{Q}\,cos(\theta_m) \, ,
\end{array}
\label{piobs}
\end{equation}
where $\theta_m \neq \theta_T$. According to Eq.~\ref{piproj} the noise-free $P_T$ is somewhat different from $P_m$: 
\begin{equation}
P_T + \delta{P} = P_m\ cos(\theta_T - \theta_m) 
\end{equation}
The assumption is that the median angle $\theta_m$ is very close to the true angle $\theta_T$,
so $\left|\delta{\theta}\right| = \left|\theta_T - \theta_m\right| \ll 1$, and $sin(\delta{\theta}) < \delta{\theta}$. \\

The accuracy of the determination of $P$ depends directly on the angle error $\delta{\theta}$. Therefore we can express the error in $P_m$ as a function of $\delta{\theta}$:
\begin{equation}
 \begin{array}{lcl}
  \delta{P(\delta{\theta)}} &=& P_m\ \frac{\partial{}}{\partial{\theta_m}} cos(\theta_T-\theta_m) \ \delta{\theta} \\
                      &=& P_m\,sin(\theta_T-\theta_m) \ \delta{\theta} \\
                      &<& P_m\,(\delta{\theta})^2 \, .
 \end{array}
 \label{eqnoise}
\end{equation}
On the other hand, the angle error $\Delta{\theta}$ generally depends on the errors of $U$ and $Q$:
\begin{equation}
 \begin{array}{lcl}
  (\Delta{\theta}(U,Q))^2 &=& [\frac{\partial{}}{\partial{U}} {tan^{-1}(U/Q)}\ \Delta{U}]^2 \\
  &+& [\frac{\partial{}}{\partial{Q}} {tan^{-1}(U/Q)}\ \Delta{Q}]^2 \\
  &=& [(Q\,\Delta{U})^2 + (U\,\Delta{Q})^2]\,/\,(U^2 + Q^2)^2 \, . \\
 \end{array}
 \label{dP_theta}
 \end{equation}
In addition, we have to take into account the contribution of the modified median filter to the angle error. If we apply an $n\times n$ filter box we get
\begin{equation}
\delta{\theta}_m(U,Q) \approx \Delta{\theta}(U, Q) / n \, .
\end{equation}
Using $sin(\delta{\theta_m}) < \delta{\theta_m}$, the error of $P_m$ with respect to the angle error depending on $\Delta{U}$ and $\Delta{Q}$ is:
\begin{equation}
 \begin{array}{lcl}
   \delta{P}(\delta{\theta_m}) &=& P_m\sin(\delta{\theta_m})\  \delta{\theta_m}\\
  &<& [(Q\,\Delta{U})^2 + (U\,\Delta{Q})^2]\,/\,[n^2\,P_m^3] \, .
 \end{array}
\end{equation}
The higher the signal-to-noise ratio, the smaller the angle error $\delta{\theta_m}$.
In the case where $\Delta{U} \approx \Delta{Q}$,
\begin{equation}
\delta{P}(\delta{\theta_m}) < (\Delta{U}\,\Delta{Q}) / (n^2 \ P_m) \, ,
\end{equation}
and we can neglect the error of $P$ with respect to the angle error $\delta{\theta_m}$ in all cases. \\

The above section explained that the contribution of $\Delta{\theta}$ to the error of $\Delta{P}$ is negligible.
Thus the error depends only on $\Delta{U}$ and $\Delta{Q}$, following Eq.~\ref{piobs}:
\begin{equation}
(\Delta{P}(U, Q))^2 = sin^2(\theta_m)\,(\Delta{U})^2 + cos^2(\theta_m)\,(\Delta{Q})^2 \, .
\end{equation}
If $U$ and $Q$ are obtained from correlated signals, then $\Delta{U} = \Delta{Q}$, and $\Delta{P} = \Delta{U} = \Delta{Q}$.

\begin{figure}[h]
 \includegraphics[scale=0.4,keepaspectratio=true,clip=true,trim=30pt 0pt 0pt 0pt]{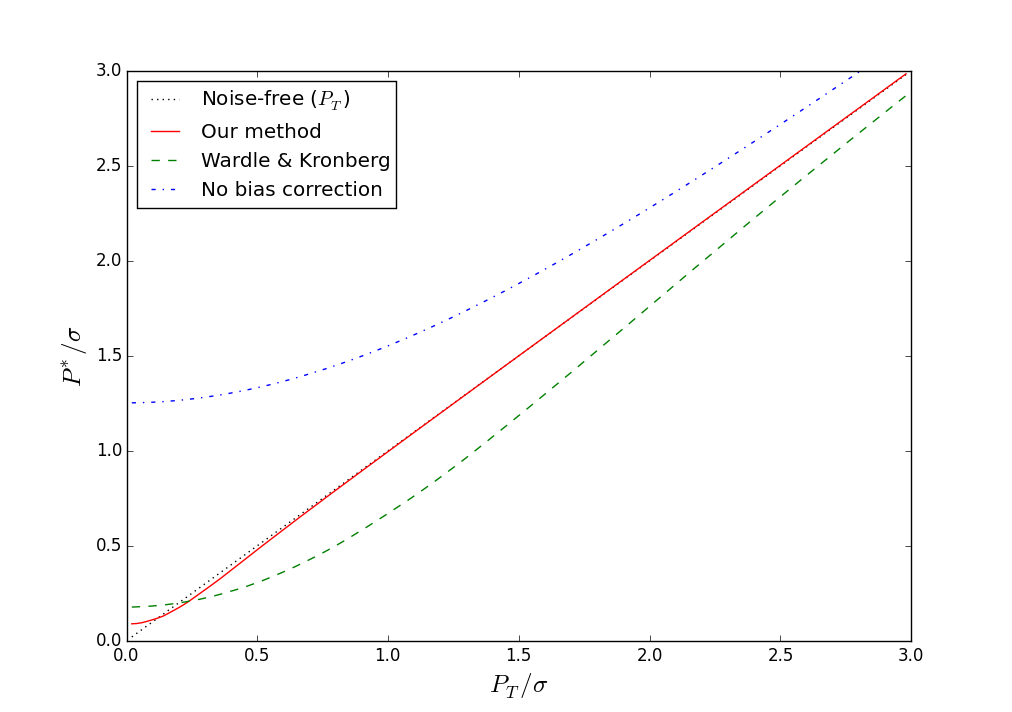}
 \caption{Averaged signal-to-noise ratio ($P^*/\sigma$) as a function of the true ratio $P_T/\sigma$ after applying: the bias suppression method; the Wardle \& Kronberg method (using $C=1.2$); our new method;
 and without any bias correction.
 To show the improvement of the bias suppression with our new method, we created a $100 \times 100$ pixel $\hat{U}$ map and $\hat{Q}$ map that contained only Gaussian
 noise. We added a constant positive value $A$ to the entire $\hat{U}$ map and a negative value $-A$ to the $\hat{Q}$ map.
 The values vary between $0<|A|<3\sigma$. For each $A$ we calculated the corresponding $P^*$ as the average value over the entire map for all methods given in the plot.
 }
 \label{bias}
\end{figure}

According to Fig.~\ref{bias} the corrected polarised intensity $P^*$ is always overestimated when no bias correction is applied, whereas the  Wardle \& Kronberg method
always underestimates $P^*$.
Our new method recovers the true signal $P$ more precisely, except for very small signal-to-noise ratios.

Our method is fundamentally different from that of Wardle \& Kronberg. 
We are able to determine the polarised intensity, including the noise, by projecting the observed values $\hat{U}$ and $\hat{Q}$ onto the direction of the median filtered
polarisation angle $\theta_m$ (Eq.~9). The noise distribution of $P$ is the same as that of $U$ und $Q$ (Fig.~7) and hence averages to zero 
over large areas. In contrast to our method, Wardle \& Kronberg use the mean noise ($rms$) of the entire map to correct $P$ for the bias.
We do not need to know the mean errors of $\hat{U}$ and $\hat{Q}$. Furthermore, the distributions of these errors do not necessarily need to be Gaussian. 

Systematic errors (in addition to pure noise), such as fluctuations of the background emission level in Stokes $U$ and $Q$ on the scales of the beam or instrumental polarisation
(usually below 1\% of the total emission), lead to additional signals in polarised intensity that cannot be corrected by the methods of bias suppression. Instrumental 
polarisation should be reduced by applying the Mueller matrix to $U$ and $Q$ before the determination of $P$.

\section{Numerical simulations}

To test our new method, simulated images are created that contain an artificial source (e.g. a galaxy) and four box
shaped structures of size $40\times33$ pixels. These simulate abrupt changes at the edges and constant polarised intensity inside
the boxes. The artificial galaxy is composed of two two-dimensional Gaussians, a circular one at the central region
and a large elliptical one simulating the halo. The simulated Stokes $U$ contains only the halo, whereas the $Q$ map
contains both a halo of the artificial galaxy component and the centre source. The boxes have different amplitudes of
$\pm 0.5\cdot{rms}$, $\pm 1.5\cdot{rms}$, $\pm 3.0\cdot{rms}$, and $\pm 5.0\cdot{rms}$, in which the component of Q
is always positive and U always negative. The amplitudes of the boxes increases from the upper left anticlockwise
to the upper right. 

\begin{figure}[h]
 \centering
 \includegraphics[scale=0.475,keepaspectratio=true]{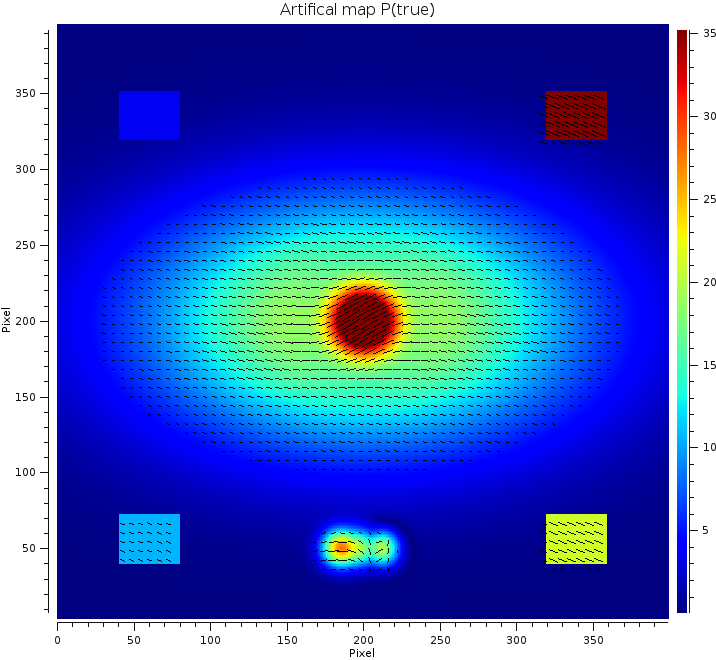}
 \caption{Polarised intensity reference map. $P$ is calculated from the noise-free $U$ and $Q$ maps using $P = \sqrt{U^2 + Q^2\ }$.
 The lines indicate the orientation of the polarised emission.}
 \label{PItrue}
 \end{figure}

To test the effect of different polarisation angles on the scale of the beam we add three overlapping Gaussian sources with amplitudes of 
$+5\cdot{rms}$, $2\cdot{rms}$, $-5\cdot{rms}$ to $Q$ and $+2\cdot{rms}$, $-5\cdot{rms}$, $+2\cdot{rms}$ to $U$. We note that in this case we get depolarisation in the 
overlapping regions of the sources. From the two maps of $U$ and $Q$ the polarised intensity is constructed as $P = \sqrt{U^2 + Q^2}$, which is the noise-free 
reference map (Fig.~\ref{PItrue}).

\begin{figure}[h]
 \centering
 \includegraphics[scale=0.475,keepaspectratio=true]{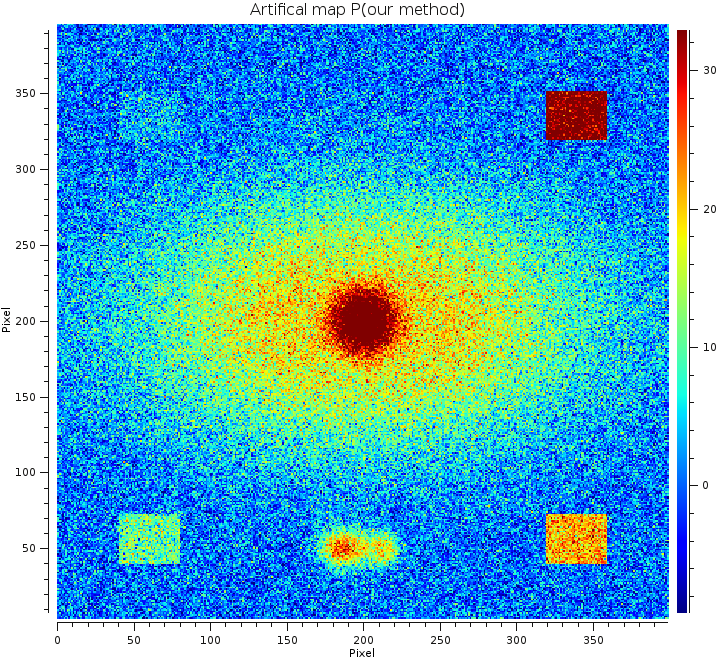}
 \caption{Polarised intensity map calculated with our new method.}
 \label{PI}
 \end{figure}

Gaussian noise $N_U$ and $N_Q$ with $rms = 5$ is added to both the $U$ and $Q$ maps. Subsequently the bias correction performed with our method by applying the
modified median filter (MMF) to the angle map $\theta_m = MMF(tan^{-1}[(U_i+N_U)\,/\,(Q_i+N_Q)])$.
Then the bias corrected polarised intensity $P^*$ is computed using Eq.~9 which is shown in Fig.~\ref{PI}.

\begin{figure}[h]
 \centering
 \includegraphics[scale=0.475,keepaspectratio=true]{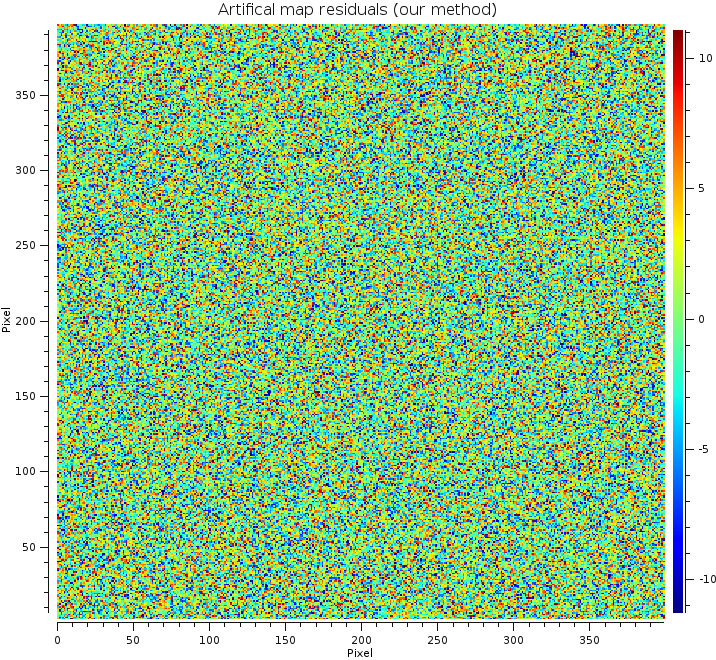}
 \caption{Residual map from our new method. Only Gaussian noise is showing after subtracting the reference map.}
 \label{diff}
 \end{figure}

 \begin{figure}[h]
 \centering
 \includegraphics[scale=0.475,keepaspectratio=true]{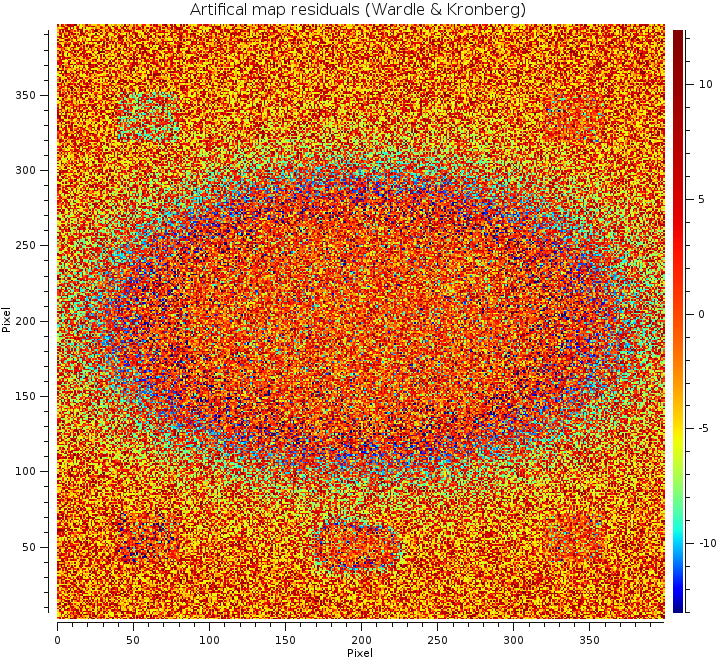}
 \caption{Residual map from the Wardle \& Kronberg method. Artefacts are showing at low intensities.}
 \label{diffOld}
 \end{figure}

\begin{figure}[h]
 \centering
 \includegraphics[scale=0.4,keepaspectratio=true,clip=true,trim=0pt 0pt 0pt 0pt]{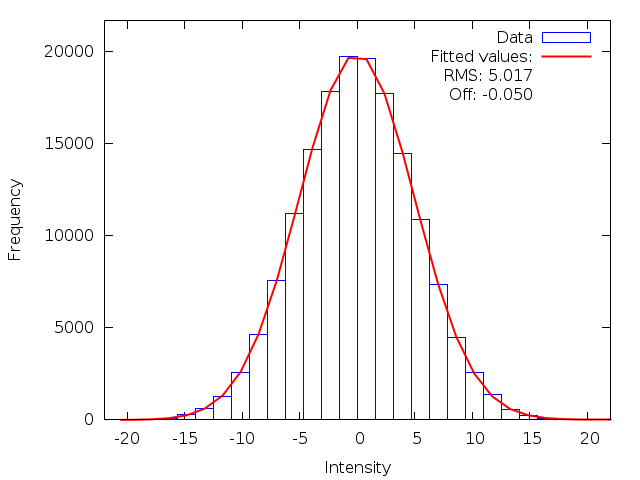}
 \caption{Histogram of the residual map from our new method (Fig.~5). Intensities are given in arbitrary numbers.}
 \label{Histo}
 \end{figure}

\begin{figure}[h]
 \centering
 \includegraphics[scale=0.4,keepaspectratio=true,clip=true,trim=0pt 0pt 0pt 0pt]{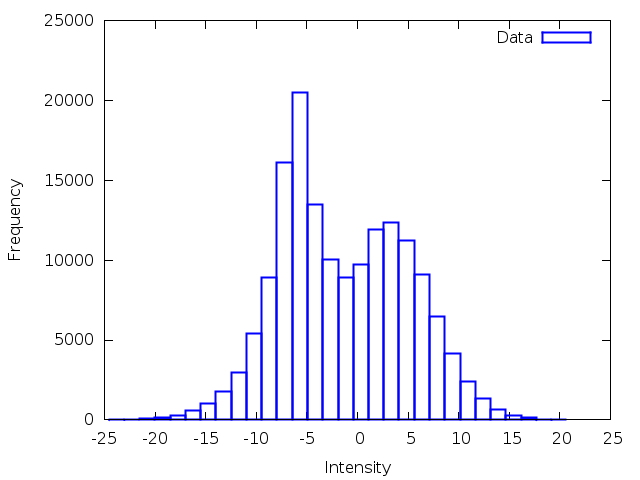}
 \caption{Histogram of the double-peaked residual map from the Wardle \& Kronberg method (Fig.~6). Intensities are given in arbitrary numbers.}
 \label{Histo_Old}
 \end{figure}

After subtracting the reference map, the residual map should contain only noise, if the method has worked properly. This is the case for our method (Fig.~\ref{diff}) 
but not for the method of Wardle \& Kronberg (Fig.~\ref{diffOld}). We note that even in the case of three overlapping sources with different 
polarisation angles our method works well, due to the behaviour of the median filter. The histogram (Fig.~\ref{Histo}) of the residual map shows Gaussian noise distribution 
with $rms = 5$ and centred close to zero. The rms is the same that was added to $U$ and $Q$, as expected. Compared with the Wardle \& Kronberg method
(Fig.~\ref{Histo_Old}) there is a significant improvement. At low intensities the Wardle \& Kronberg method shows artefacts which are not seen in our new method.

Table~1 shows the results of the two methods, comparing the total integral of the entire maps. The relative error of our new method is significantly smaller than the
relative error of the Wardle \& Kronberg method. Also the relative errors of the boxes of our new method are smaller than those of the
Wardle \& Kronberg method (Table~2).

\begin{table}  
\begin{tabular}{|l@{\hspace{1cm}}|@{\hspace{1cm}}c|@{\hspace{1cm}}c|}
 \hline
 \ & Sum & Error\\
 \hline
 P(true) & 871808 &  - \\
 \hline
 P(new) & 876685 &  0.6\% \\
 \hline
 P(old) & 793768 &  9.0\% \\
 \hline
  P(bias) & 1471320 & 68.8\% \\
 \hline
\end{tabular}
\caption {Sum of polarised intensities of the simulated images over all pixels. P(bias) is taken without any bias correction.}
\end{table}

\begin{table}
  \scalebox{0.825}{%
  \begin{tabular}{|l|l|l|l|l|l|l|l|l|}
    \hline
    \multirow{2}{*}{ \ } &
      \multicolumn{2}{c|}{Box1} &
      \multicolumn{2}{c|}{Box2} &
      \multicolumn{2}{c|}{Box3} &
      \multicolumn{2}{c|}{Box4} \\
    & Sum & Error & Sum & Error & Sum & Error & Sum & Error \\
    \hline
    P(true) & 4412 & - & 13237 & - & 26474 & - & 44124 & - \\
    \hline
    P(new) & 4086 & \ 7.4\% & 13522 & 2.2\% & 26383 & 0.3\% & 44074 & 0.1\% \\
    \hline
    P(old) & 2283 & 48.3\% & 12044 & 9.0\% & 26011 & 1.7\% & 43919 & 0.5\% \\
    \hline
    P(bias) & 8553 & 93.8\% & 15037 & 13.6\% & 27145 & 2.5\% & 44565 & 1.0\% \\
    \hline
  \end{tabular}}
  \caption {Sum of polarised intensities of the simulated images over each box.}
\end{table}

\section{Application to real data: M31}

The radio data of the Andromeda galaxy M31 (Fig.~9), observed with the Effelsberg 100-m telescope at 4.85\,GHz (Berkhuijsen et al. 2003), can be used to effectively 
demonstrate the advantages of our new method. The maps have $3\arcmin$ resolution and a pixel size of $1\arcmin$.
The polarisation map determined with our new method (Fig.~10) agrees with that obtained with the Wardle \& Kronberg method (Fig.~11) at medium and high 
intensities, but the maps differ significantly at low intensities.
The histogram for the new map is smooth (Fig.~12), whereas the histogram of the map determined with the Wardle \& Kronberg 
method (Fig.~12) shows a sharp minimum at $P^*=0$ and an increase towards negative values of
$P^*$, which is the reason for the deep minima in Fig.~11. The lowest value in the $P^*$ map created with our new method (Fig.~10) is
-740\,mJy/beam, while the map using the Wardle \& Kronberg method is limited to about -293\,mJy/beam.
Using our new method, the integrated polarised flux density is 14\% larger than that of the Wardle \& Kronberg method.

\begin{figure}[h]
 \centering
 \includegraphics[scale=0.425,keepaspectratio=true,clip=true,trim=50pt 0pt 0pt 0pt]{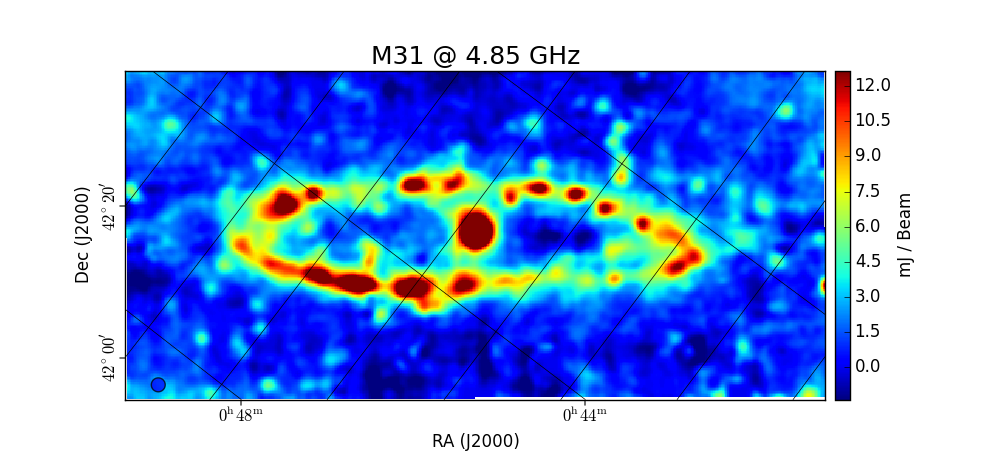}
 \caption{Total intensity of M31 at 4.85\,GHz (in mJy/beam), observed with the Effelsberg 100-m telescope at an angular resolution of $3\arcmin$ and $1\arcmin$ pixel size. 
 Bright background sources have been subtracted (from Berkhuijsen et al. 2003).}
 \label{m31_tp}
\end{figure}

\begin{figure}[h]
 \centering
 \includegraphics[scale=0.425,keepaspectratio=true,clip=true,trim=50pt 0pt 0pt 0pt]{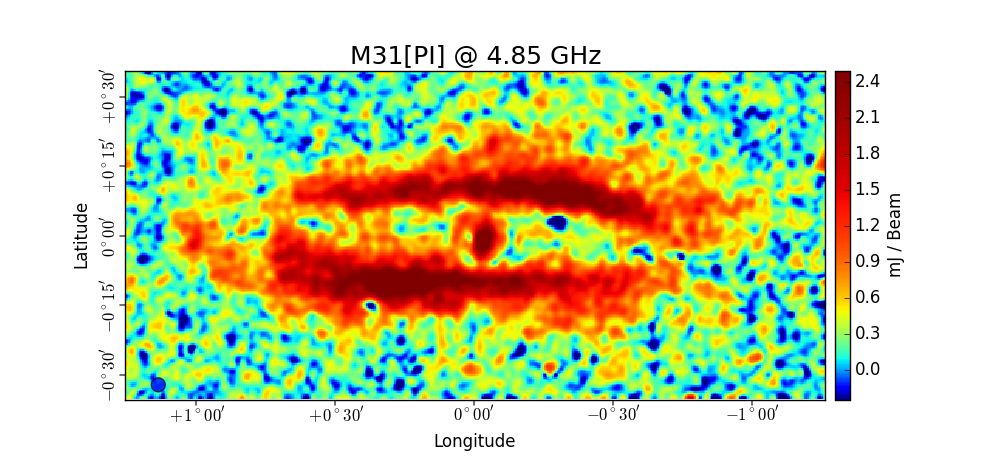}
 \caption{Polarised intensity of M31 at 4.85\,GHz (in mJy/beam) determined with our method and an angular
 resolution of $3\arcmin$.}
\end{figure}

\begin{figure}[h]
 \centering
 \includegraphics[scale=0.425,keepaspectratio=true,clip=true,trim=50pt 0pt 0pt 0pt]{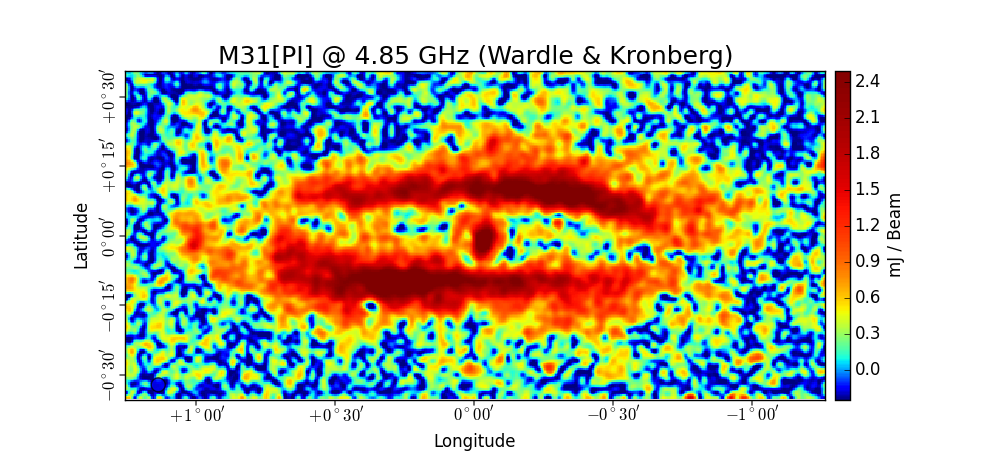}
 \caption{Polarised intensity of M31 at 4.85\,GHz (in mJy/beam) determined with the
 Wardle \& Kronberg method, assuming a 0.2\,mJy/beam rms noise for $U$ and $Q$.}
 \label{oldMethod}
\end{figure}

\begin{figure}[h]
 \centering
 \includegraphics[scale=0.4,keepaspectratio=true,clip=true,trim=0pt 0pt 0pt 0pt]{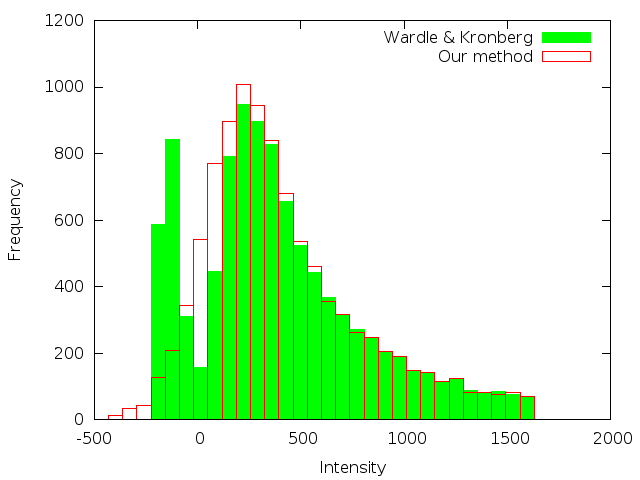}
 \caption{Histograms of polarised intensity of M31 at 4.85~GHz: our new method (red) and the Wardle \& Kronberg method (green).}
 \label{histWD}
\end{figure}

The radio polarisation data of the Andromeda galaxy M31 observed with the Very Large Array (VLA) at 1.4\,GHz (Beck et al. 1998) reveals 'depolarisation canals' at locations
where the values of $\hat{U}$ and $\hat{Q}$ both cross the zero level and the polarisation angle jumps by $90\degr$ across the canals. The width is observed to be
smaller than the telescope beamwidth, which needs steep gradients in $\hat{U}$ and $\hat{Q}$. Such canals could be produced either by differential Faraday rotation 
(Shukurov \& Berkhuijsen 2003) or by beam depolarisation in a turbulent magneto-ionic medium (Haverkorn \& Heitsch 2004). The length, depth, and separation of the 
canals provide tools for studies of interstellar turbulence.
The widths of the canals at half intensity, $60\arcsec - 100\arcsec$ in the map obtained with the Wardle \& Kronberg method
(Fig.~14), are $20\% - 30\%$ larger in the new map (Fig.~13). This relaxes the need for steep gradients in Faraday rotation measures and hence reduces the 
amplitude needed for studies of interstellar turbulence. The difference between the maps of Figs.~13 and 14 is particularly significant in the canals and other regions of low intensity (Fig.~15). 
 The canals are less deep when applying our method.

\begin{figure}[h]
 \centering
 \includegraphics[scale=0.425,keepaspectratio=true,clip=true,trim=50pt 0pt 0pt 0pt]{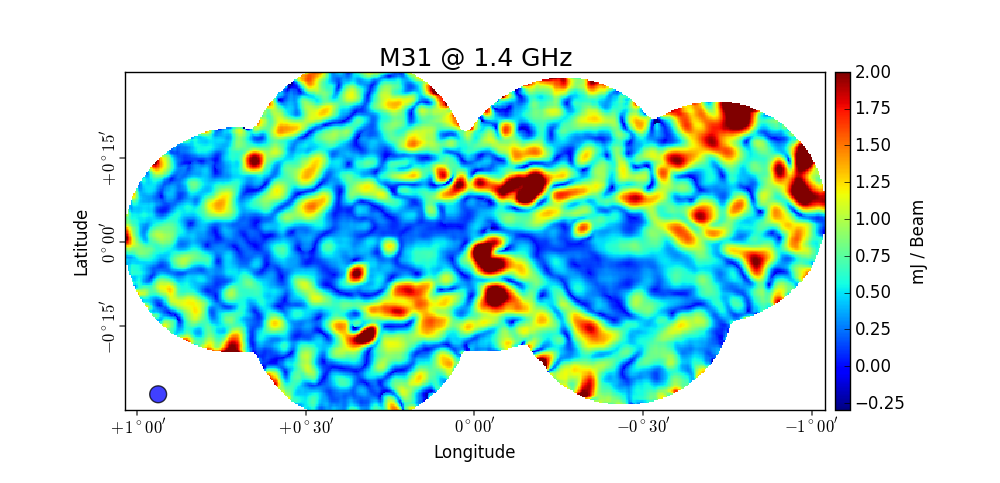}
 \caption{Polarised intensity of M31 at 1.4\,GHz (in mJy/beam) determined with our method and an angular
 resolution of $3\arcmin$.}
\end{figure}

\begin{figure}[h]
 \centering
 \includegraphics[scale=0.425,keepaspectratio=true,clip=true,trim=50pt 0pt 0pt 0pt]{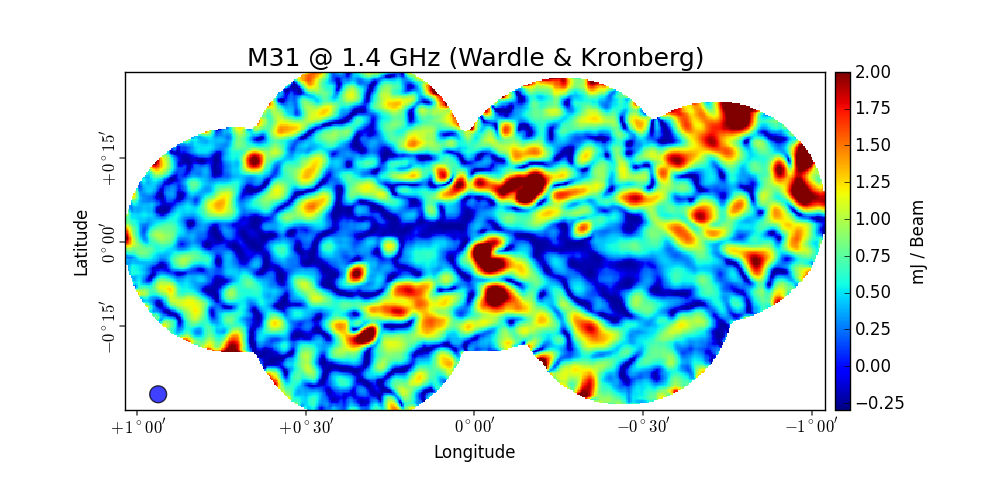}
 \caption{Polarised intensity of M31 at 1.4\,GHz (in mJy/beam) determined using the
 Wardle \& Kronberg method and assuming an rms noise of 0.2\,mJy/beam.}
\end{figure}

\begin{figure}[h]
 \centering
 \includegraphics[scale=0.425,keepaspectratio=true,clip=true,trim=50pt 0pt 0pt 0pt]{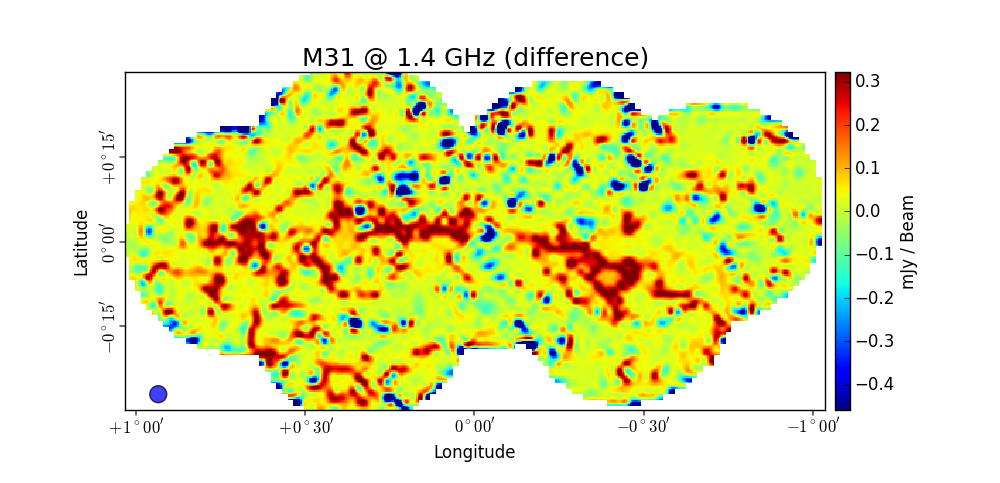}
 \caption{Difference map of our method (Fig.~13) and the Wardle \& Kronberg method (Fig.~14).}
\end{figure}

Equation~7 describes the projection of $N_U$ and $N_Q$ to $N_P$. We then show that the noise distribution of $N_P$ has the same characteristics as $N_U$ and
$N_Q$. Therefore any other projections of two orthogonal noise components that result in $N_P$ also have the same characteristics. The projection described by
Eq.~8 can be used to estimate the noise contribution $N_{P_m}$. The noise biased $\hat{P}$ is the vector addition of the true polarised intensity $P_T$ and the
noise contribution $N_{\hat{P}}$ of $\hat{U}$ and $\hat{Q}$. $N_{\hat{P}}$ is also projected onto the direction of $P_T \approx P_m$. Although the cosine component
$P^* = P_m + N_{P_m}$ does not allow us to separate the noise contribution, the sine component $N_P'$ is due to pure noise:
\begin{equation}
 N_P' = \hat{P}\ sin(\Delta{\Theta})\, ,  \ \ $where$ \ \  \Delta{\Theta} = \theta_m - \hat{\theta} \, , \\
\label{error}
\end{equation}
which is shown in Fig. 16.
This equation can be used as an estimate of the noise distribution of the $P^*$ map created with our method. It does, however, not include systematic errors arising from 
deviations between $\theta_m$ and $\theta_T$. These become relevant for very small signal-to-noise ratios (the remaining bias is shown in Fig. 2) and in cases
of strong small-scale variations of $\theta$, in which the modified median cannot be a good representation of $\theta_m$. In the former case, the
statistics according to Eq.~17 still represent the noise well but do not represent residual systematic errors.

The noise components $\hat{P_m}\ cos(\Delta{\Theta}) - P_m$ and $\hat{P}\ sin(\Delta{\Theta})$ of $N_{\hat{P}}$ (as shown in Fig.~16) have the same statistical characteristics
as the noise components $N_U$ and $N_Q$ of the same vector $N_{\hat{P}}$. Statistically, both noise components are equally distributed.

\begin{figure}[h]
 \centering
 \includegraphics[scale=0.4,keepaspectratio=true,clip=true,trim=0pt 0pt 0pt 0pt]{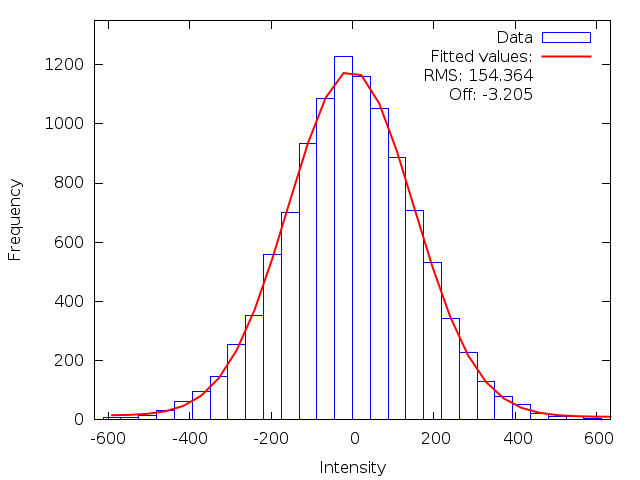}
 \caption{Histogram of the probability density of the noise distribution taken from Eq.~17.}
 \label{Hist_fromPI}
\end{figure}

This enables us to determine the probability density of the noise from the error of the angles between the
observed angle $\hat{\theta} = tan^{-1}(\hat{U} / \hat{Q})$ and the mean angle $\theta_m$,  which should be
close to the true angle $\theta_T$. Figure~\ref{Hist_fromPI} shows that the noise distribution is almost Gaussian.
The half-power width of about 157\,mJy/beam is a reliable measure of the rms noise $\sigma$ in the $P^*$ map.

\section{Conclusions}

Our simulated data show that our new method to suppress the polarisation bias almost perfectly reproduces the true
polarised intensity even in regions of very low signal-to noise ratios, which was not possible before with any of the
previously used methods. In particular, the achievements of our new method are:

\begin{itemize}

\item It estimates the polarisation angle of the source signal in a noisy environment with help of a modified
median filter. The corrected polarised intensities $P^*$ do not suffer from a general residual bias as is the case for all other
methods.
\item It works best for smooth variations of the polarisation angle in the source. In the case of
sharp jumps in polarisation angle, the modified median filter slightly increases the statistical angle error.
\item It delivers a reliable value for the rms noise. If the noise distributions in the maps of Stokes $\hat{U}$ and $\hat{Q}$ are Gaussian, the noise distribution
in the corrected $P^*$ map is also Gaussian.
\item The signal-to-noise ratios are measured directly from the $P^*$ map, without using the maps of
$\hat{U}$ and $\hat{Q}$.
\item It can also be applied if the noise distributions in the maps of Stokes $\hat{U}$ and $\hat{Q}$ are
different and/or if the distributions are not Gaussian.
\item The maps of corrected polarised intensities $P^*$ and polarisation angles are reliable even in regions with
weak signals and hence allow us to analyse the distribution of polarised intensities and polarisation angles in
faint sources. The maps are free of the artefacts produced by the Wardle \& Kronberg method (1974).
\item The corrected intensities of the polarised emission $P^*$ provide reliable integrated flux densities and degrees of polarisation
without a cumulative effect of the bias, especially for faint sources.
\item The $P^*$ map has the same noise distribution as $U$ and $Q$, which allows us to directly convolve $P^*$ signals to a larger beamsize and
hence to increase the signal-to-noise ratio for diffuse extended emission. However, caution is needed for very small signal-to-noise ratios $s$, for example when $s<0.5$.
\item Features at low intensity levels, like 'depolarisation canals', are smoother in our maps than in those produced by using the
previous methods.

\end{itemize}

\begin{acknowledgements}

We thank Axel Jessner and Aritra Basu for useful discussions and Olaf Wucknitz for critical comments on the manuscript.

\end{acknowledgements}

\end{document}